\documentclass[a4paper,fleqn,usenatbib]{mnras}

\usepackage[T1]{fontenc}
\usepackage{ae,aecompl}

\usepackage{graphicx}	
\usepackage{amsmath}	
\usepackage{amssymb}	

\def\noao{The NOAO Deep Wide Field Survey MOSAIC Data Reductions, 
http://www.noao.edu/noao/noaodeep/ReductionOpt/frames.html}%
\def\prepare{in preparation.}%

\title[Age spread in the young LMC cluster NGC\,1971]{Observational hints of a real age spread in the
young LMC star cluster NGC\,1971}

\author[A.E. Piatti \& A. Cole]{
Andr\'es E. Piatti$^{1,2}$\thanks{E-mail: andres@oac.unc.edu.ar} and Andrew Cole$^{3}$
\\
$^{1}$Consejo Nacional de Investigaciones Cient\'{\i}ficas y T\'ecnicas, Av. Rivadavia 1917, 
C1033AAJ, Buenos Aires, Argentina\\
$^{2}$Observatorio Astron\'omico, Universidad Nacional de C\'ordoba, Laprida 854, 5000, 
C\'ordoba, Argentina\\
$^{3}$School of Physical Sciences, University of Tasmania, Private Bag 37, Hobart, 7001 TAS, Australia\\
}

\date{Accepted XXX. Received YYY; in original form ZZZ}

\pubyear{2017}

\begin{document}
\label{firstpage}
\pagerange{\pageref{firstpage}--\pageref{lastpage}}
\maketitle

\begin{abstract}
We report the serendipitous young Large Magellanic Cloud 
cluster, NGC\,1971,
exhibits an extended main-sequence turnoff (eMSTO) possibly originated by mostly a real
age spread. We used $CT_1$ Washington photometry to produce a
colour-magnitude diagram (CMD) with the fiducial cluster features. From its eMSTO,
we estimated an age spread of $\sim$ 170 Myr (observed age range 100-280 Myr), 
once observational errors,
stellar binarity, overall metalicity variations and stellar rotation effects were
subtracted in quadrature from the observed age width. 

\end{abstract}




\section{Introduction}

A recent study on the population of extended main-sequence turn-off (eMSTO) clusters
in the Large Magellanic Cloud (LMC) showed that the eMSTO phenomenon is not caused by actual age spread
within the cluster \citep{pb16b}. Furthermore, the authors confirmed 
that clusters with log($t$ yr$^{-1}$) $\sim$ 9.2$-$9.3 have broader MSTOs than younger 
and older clusters on average, in very good agreement with \citet[][see their Fig. 4]{bastianetal16}.
For clusters younger than 1 Gyr they showed a strong correlation with age as found by
\citet{niederhoferetal15b} and discussed at length in \citet{niederhoferetal16}.
The latter invoked stellar evolutionary effects (e.g. due to stellar rotation) to
explain the eMSTOs, which may mimic the effect of an age spread.

The way to test if age is the defining parameter of eMSTOs is to look for them
in younger clusters ($<$500 Myr). Indeed, NGC\,1755 \citep[][$\sim80$~Myr]{miloneetal16},
NGC\,1850 \citep[][$\sim100$~Myr]{bastianetal16}, 
NGC\,1856 \citep[][$\sim300$~Myr]{dantonaetal15,lietal2017}, show small eMSTOs, consistent 
with the trend showed by \citet{niederhoferetal15b}. 

In this Letter, we report the first young LMC cluster, NGC\,1971, with an observed 
real age spread as shown by the Washington photometry colour-magnitude diagram (CMD)
analysis. The cluster, located in the South-Eastern half of the LMC bar, has only been 
photometrically studied in detail by \citet{dg2000}, so that 
there was no clue about a possible 
unusual broadness at its MSTO as to pay attention to.  This finding points to the need for
further observations in order to investigate in detail the eMSTO phenomenon in NGC\,1971, 
namely, the fraction of blue and red MS stars, their spatial radial distributions, as well
as spectrocopic follow-ups to tag metal abundance inhomogeneities and kinematics.
In Section 2 we describe the observational data set, while Section 3 deals with the
construction of the cluster CMD. Section
4 presents the estimation of its astrophysical properties and the analysis about the possible 
origin of the observed eMSTO.

\section{Observational data}

\begin{figure}
	\includegraphics[width=\columnwidth]{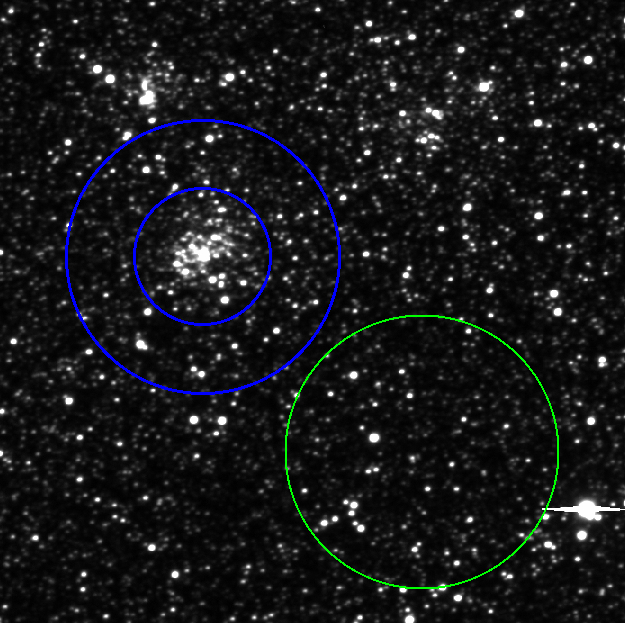}
    \caption{3$\times$3 arcmin$^2$ $R$ image centred on the NGC\,1971 field. North is up and East to
the left. Blue circles of 20 and 40 arcsec and a green circle of 40 arcsec are shown. The cluster
region up to 40 arcsec was consider for the completeness analyses, but stars within 20 arcsec are used for 
the CMD. NGC\,1969 and NGC\,1972
are seen to the North and North-West of NGC\,1971, respectively.}
   \label{fig:fig1}
\end{figure}

We made use of $CT_1$ Washington images available at the National Optical Astronomy Observatory (NOAO) Science Data Management (SDM) 
Archives\footnote{http://www.noao.edu/sdm/archives.php.}.  They were obtained at the 
Cerro-Tololo Inter-American Observatory (CTIO) 4 m Blanco telescope with the Mosaic \,II 
camera attached (36$\arcmin$$\times$36$\arcmin$ field onto a 8K$\times$8K CCD detector array) 
as part of a survey of the most metal-poor stars outside the Milky Way 
(CTIO 2008B-0296 programme, PI: Cole). The images analyzed here consist of a 420 s $C$ and 
a 30 s $R$ exposures carried out under photometric conditions at an airmass of 1.3. 
Fig.~\ref{fig:fig1} shows an enlargement of the $R$ image centred on the cluster region.

We processed the images, measured 
the instrumental magnitudes and stardardized the photometry in a comprehensive way, as previously
performed for similar data sets \citep [e.g.][and references therein]{pietal12,p12a,p15}, together
with the whole data set for the aforementioned CTIO programme, which comprises 17 different
LMC fields, and 3 standard fields observed three times per night (Dec.27-30, 2008) to secure the quality of the transformation equations \citep{petal2017}.
We summarize here some specific issues in order to provide the reader with an overview
of the photometry quality. The data reduction followed the procedures documented by the NOAO Deep Wide 
Field Survey team \citep{jetal03}
and utilized the {\sc mscred} package in IRAF\footnote{IRAF is distributed by the National 
Optical Astronomy Observatories, which is operated by the Association of 
Universities for Research in Astronomy, Inc., under contract with the National 
Science Foundation.}. We performed overscan, trimming and cross-talk corrections, bias subtraction,
flattened all data images, etc., once the calibration frames (zeros, sky- and dome- flats, etc) were properly 
combined. For each image we obtained
 an updated world coordinate system (WCS) database with a rms error in right ascension and 
declination smaller than 0.4 arcsec, by using $\sim$ 500 stars catalogued by the 
USNO\footnote{http://www.usno.navy.mil/USNO/astrometry/optical-IR-prod/icas/usno-icas}. 

The stellar photometry was performed using the star-finding and point-spread-function (PSF) fitting 
routines in the {\sc daophot/allstar} suite of programs \citep{setal90}. We measured magnitudes on  the single 
image created by joining all 8 chips together using the updated WCS. This allowed us to use a unique 
reference coordinate system
for each LMC field. For each Mosaic image, a quadratically varying 
PSF was derived by fitting $\sim$ 1000 stars (nearly 110-130 stars per chip), once the neighbours
 were eliminated using a preliminary PSF
derived from the brightest, least contaminated $\sim$ 250 stars (nearly 30-40 stars per chip). Both
 groups of PSF 
stars were interactively selected. We then used the {\sc allstar} program to apply the resulting
 PSF to the 
identified stellar objects and to create a subtracted image which was used to find and measure
 magnitudes of 
additional fainter stars. This procedure was repeated three times for each frame. 
We computed aperture corrections from the 
comparison of PSF and
aperture magnitudes by using the neighbour-subtracted PSF star sample.
Finally, we standardized
the resulting instrumental magnitudes and combined all the independent measurements using the
 stand-alone {\sc daomatch} 
and {\sc daomaster} programs\footnote{Provided kindly by Peter Stetson.}.  
Note that the $R$ filter has significantly 
higher throughput as compared with the standard Washington 
$T_1$ filter so that $R$ magnitude can be accurately transformed to yield $T_1$ 
magnitudes \citep{g96}.

We first examined the quality of our photometry in order to evaluate the influence of the photometric 
errors, crowding effects and the detection limit on the cluster CMD fiducial characteristics. 
To do this, we performed artificial star tests to derive the completeness
level at different magnitudes. We used the stand-alone {\sc addstar} program in the {\sc daophot}
package \cite{setal90} to add synthetic stars, 
generated bearing in mind the colour and magnitude distributions 
of the stars in the CMDs (mainly along the cluster main-sequence), as well as its radial stellar 
density profile. We added a number
of stars equivalent to $\sim$ 5$\%$ of the measured stars in order to avoid in the synthetic images 
significantly 
more crowding than in the original images. On the other hand, to avoid small number statistics in the
 artificial-star 
analysis, 
we created a thousand different images for each original one. We used the option of entering the number of
 photons
per ADU in order to properly add the Poisson noise to the star images. 

\begin{figure}
	\includegraphics[width=\columnwidth]{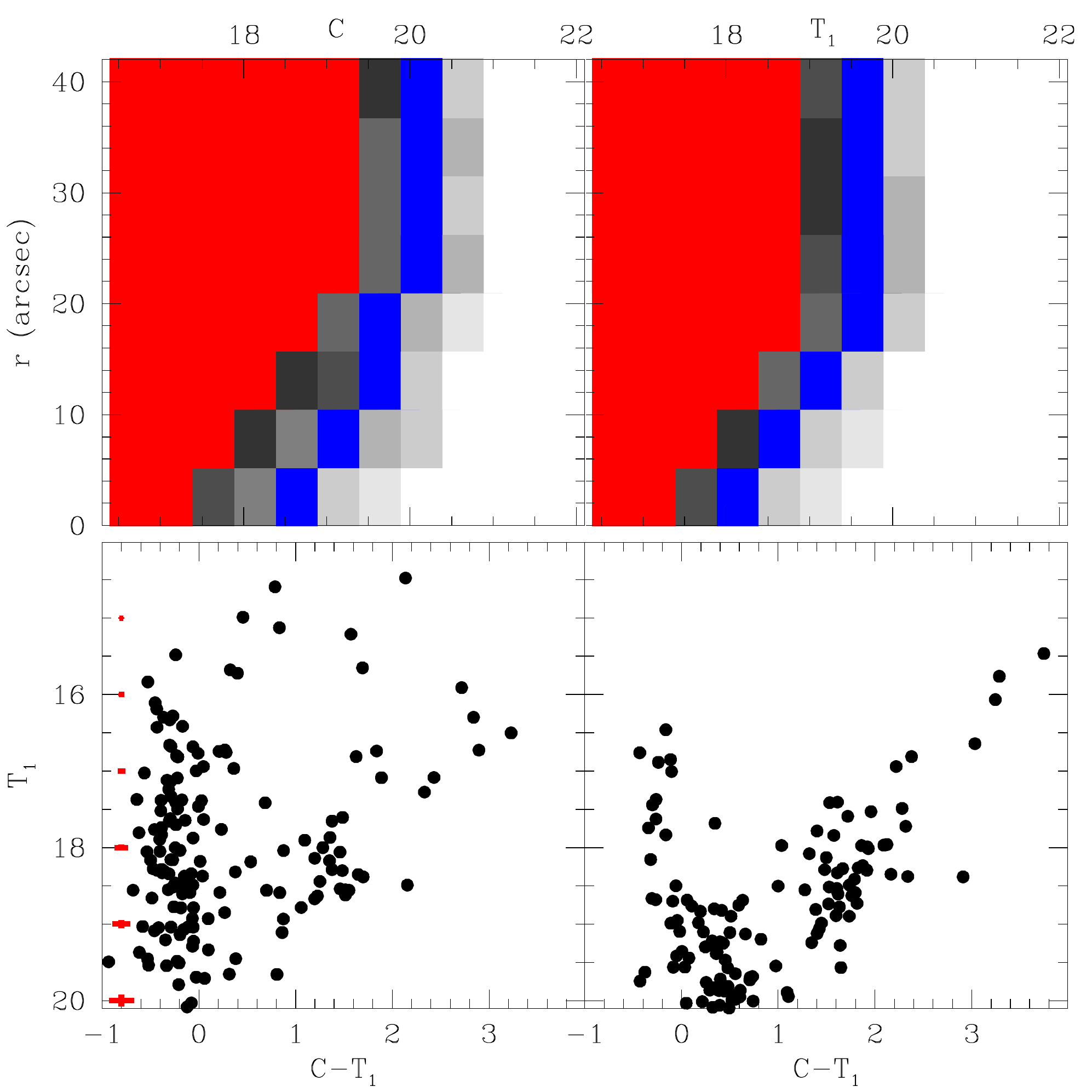}
    \caption{{\it Top:} Gray-scale completeness fraction for the NGC\,1971
field. The higher the completeness fraction the darker the symbol.
Red and blue regions represent the 100\% and 50\% completeness
levels, respectively. {\it Bottom:}  CMD for all the stars measured within 20 
arcsec of NGC\,1971, with errorbars obtained from 
artificial star tests drawn at  the left margin (left panel),
 and for a star field located adjacent to 
NGC\,1971, for the same cluster area (right panel).
}
   \label{fig:fig2}
\end{figure}

We then repeated the same steps to obtain the photometry of the synthetic images as described above, 
i.e., 
performing three passes with the {\sc daophot/allstar} routines. 
The star-finding efficiency were estimated by comparing the output 
and the input data for these stars using the {\sc daomatch} and {\sc daomaster} tasks.
We illustrate in Fig.~\ref{fig:fig2} the resultant completeness fractions in the radius versus
magnitude plane. 
As can be seen, the dependence of the completeness fraction with the distance from the cluster
centre nearly starts at a radius $r_{cls}$ $\sim$ 20 arcsec, which corresponds to the distance from the cluster centre 
where the combined cluster plus background stellar 
density profile is no longer readily distinguished from a constant background value  within 1-$\sigma$
of its fluctuation. We adopt this radius to build the cluster CMD.
Note that we are not interested in properties such as the cluster's structure or stellar density
profile, but in the stars which allow us to meaningfully
define the cluster's fiducial sequences in its CMD. 

\section{NGC\, 1971's CMD}

The CMD for stars with $T_1, C-T_1$ measurements located within the adopted cluster radius
is shown in Fig.~\ref{fig:fig2}. 
To filter the field stars from the CMD, we applied a statistical procedure developed by
\citet{pb12} and successfully used elsewhere 
\citep[e.g.][and references therein]{p14,petal15a,petal15b,pb16a}, which consists, 
firstly, in adopting four CMDs from different regions located reasonably far from the
cluster, but not too far so as to risk losing the local field-star
signature in terms of stellar density, luminosity function and/or colour distribution. 
Each field, with area 2$\pi$$r_{cls}$, was used as a reference to statistically filter the stars in an
equal circular area centred on NGC\,1971 (see, as an example, 
the green circle in Fig.~\ref{fig:fig1}). 

Starting with reasonably large boxes -- typically 
($\Delta$($T_1$),$\Delta$($C-T_1$)) = (1.00, 0.50) mag -- centred on each
star in the four field CMDs and by subsequently reducing their sizes
until they reach the stars closest to the boxes' centres in magnitude and colour,
separately, we defined boxes which result in use of larger areas in
field CMD regions containing a small number of stars, and vice versa.
Note that the definition of the position and size of each box involves
two field stars, one at the centre of the box and another -the closest one to 
box centre - placed on the boundary of that box. \citet{pb12} have shown that this is
an effective way of accounting for the local field-star
signature in terms of stellar density, luminosity function and/or
colour distribution.

\begin{figure*}
	\includegraphics[width=15cm]{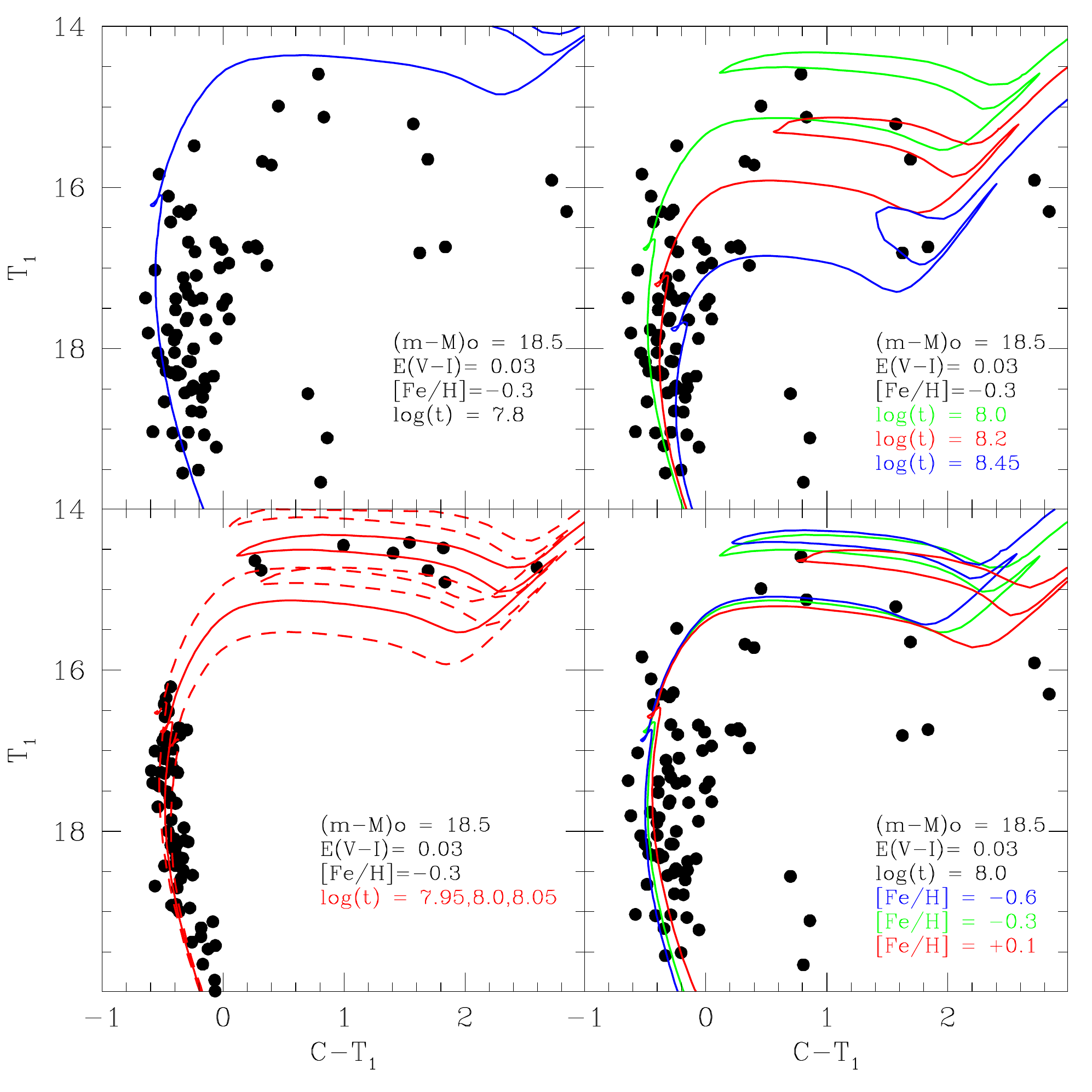}
    \caption{Cluster CMD for stars with membership probabilities
$P \ge$ 75\% with different isochrones superimposed as indicated
in each panel. A synthetic cluster CMD with is shown in the
bottom-left panel.}
   \label{fig:fig3}
\end{figure*}

When comparing the four cleaned CMDs, we counted the number of times a star
remained unsubtracted in all of them. Thus, we distinguished stars that
appear once, twice, until four times, respectively. Stars appearing
once can be associated to a probability $P \le 25\%$ of being a fiducial
feature in the cleaned CMD, i.e., stars that could most frequently be found
in a field-star CMD. Stars that appear twice ($P = 50\%$) could equally 
likely be associated with either the field or the object of interest; and
those with $P \ge$ 75\%, i.e., stars found in three or four cleaned CMDs,
belong predominantly to the cleaned CMD rather than to the field-star CMDs. 
Statistically speaking, a certain amount of residuals is expected, 
which depends on the degree of variability of the stellar density,
luminosity function and colour distribution of the star fields.
Fig.~\ref{fig:fig3} (upper-left panel) depicts the resulting CMD for stars
that comply with $P \ge$ 75\%. Since the stars are concentrated in a very
small area, we are confident of dealing with fiducial cluster features.
In the figure, we superimposed the isochrone computed by \citet{betal12} 
corresponding to the solution 
adopted by \citet{dg2000} for the cluster properties, i.e., $E(V-I)$ = 0.03 mag,
$(m-M)_o$ = 18.5 mag, Z=0.008 dex and log($t$ yr$^{-1}$) = 7.8.

\section{Analysis and discussion}

As can be seen (Fig.~\ref{fig:fig3}, upper-left panel), the results obtained by 
\citet{dg2000} are not a perfect match to the cluster CMD. A slighlty older isochrone
(log($t$ yr$^{-1}$) = 8.0) apparently reproduces more robustly the cluster sequences,
as is shown by the lime-coloured isochrone drawn in the upper-right panel.
Nevertheless, there still exists a large amount of stars, redder than the
isochrone colours, that make the overall fit visibly unsatisfactory. 
If we thought in a
cluster formed from a single stellar populataion, we possibly should adopt an
isochrone that runs like the one drawn in red  (log($t$ yr$^{-1}$) = 8.2) with
a mean dispersion of $\Delta$(log($t$ yr$^{-1}$)) = $^{-0.20}_{+0.25}$.
This seems to be too large an age spread to eliminate by appealing to differential
reddening, binarity, evolutionary effects (i.e. stellar rotation), and/or metallicity variations.

As far as we are aware, the reddening across the small cluster area ($r_{cls}$ 
$\sim$ 20 arcsec) is very low \citep[($E(V-I)$ = 0.03 mag][]{dg2000},
as can also be confirmed from the relatively narrow MS of the adjacent star field
shown in Fig.~\ref{fig:fig2}. On the other hand, a differential
reddening of $\Delta$$E(B-V)$ $\sim$ 0.4 mag would be needed to explain the broadness
of the upper MSTO, in constrast with very low reddening found in the LMC bar region
\citep{hetal11}.

In order to estimate the impact of photometric errors and stellar binarity on the
cluster upper MS, we generated a synthetic CMD by employig the \texttt{ASteCA} suit of functions \citep{pvp15}. We used an isochrone of log($t$ yr$^{-1}$) = 8.0 and the
above distance modulus, reddening and global metal content.
The steps by which the synthetic cluster was generated is as follows: i) the 
theoretical isochrone was densely interpolated to contain a thousand points throughout its entire length, including the most evolved stellar phases. ii) The isochrone
was shifted in colour and
magnitude according to the $E(V-I)$ and $(m-M)_o$ values to emulate the effects these extrinsic 
parameters have over the isochrone in the CMD. At this stage the synthetic cluster 
can be objectively identified as a unique point in the 4-dimensional space of 
parameters ($E(V-I)$, $(m-M)_o$, age and metallicity). iii) The isochrone was trimmed down to a certain faintest magnitude 
according to the limiting magnitude thought to be reached. iv) An initial mass function 
\citep[IMF,][]{Chabrier_2001} was sampled in the mass range $[{\sim}0.01{-}100]\,M_{\odot}$ up
to a total mass value $M_{total}$ provided via an input data file that
ensures the evolved CMD regions result properly populated.
The distribution of masses was then used to obtain a properly populated synthetic 
cluster by keeping one star in the interpolated 
isochrone for each mass value in the distribution. v) A relatively high fraction 
of stars were assumed to be binaries, which was set to  
$50\%$ \citep{miloneetal16,pb16b}, with secondary masses 
drawn from a uniform distribution between the mass of the primary star and a fraction of
$0.7$ of it. 
vi) The appropriate
magnitude completeness and photometric errors (see Section 2) were
finally applied to the synthetic cluster. 
The resulting synthetic CMD is shown in the bottom-left panel of Fig.~\ref{fig:fig3}
to which we superimpossed the isochrones of log($t$ yr$^{-1}$) = 7.95, 8.0 and 8.05, 
respectively. As can be seen, the apparent spread produced by the photometric errors
and stellar binarity, $\Delta$(log($t$ yr$^{-1}$)) = 0.05, does not explain the observed
eMSTO.

Metallicity variations were tested by matching three isochrones for the same age and
differente overall metalicities. The range of [Fe/H] values 
was adopted on the basis of the known metallicity spread of young LMC stellar populations
\citep[see, e.g.][and reference therein]{dg2000, pg13}, among others. Consequently, we 
chose as
a lower and upper limits, [Fe/H] = -0.6 dex and +0.1 dex, respectively. 
We used the isochrones of \citet{betal12}, which assume Y= 0.2485 + 1.78Z. 
The results of the isochrone matching are shown in the bottom-right panel of
Fig.~\ref{fig:fig3}, which confirms that overall metallicity inhomegeneities are neither
responsable for the observed eMSTO. We did not explore He or light elements variations, 
because we do not have available theoretical tracks for the Washington system to carry 
out such an analysis (see., e.g., Padova\footnote{http://stev.oapd.inaf.it/cgi-bin/cmd}, 
BasTI\footnote{http://www.oa-teramo.inaf.it/BASTI}, Geneva\footnote{http://obswww.unige.ch/Recherche/evoldb/index/}, Darthmouth\footnote{http://stellar.dartmouth.edu/~models/} repositories). 

\begin{figure}
	\includegraphics[width=\columnwidth]{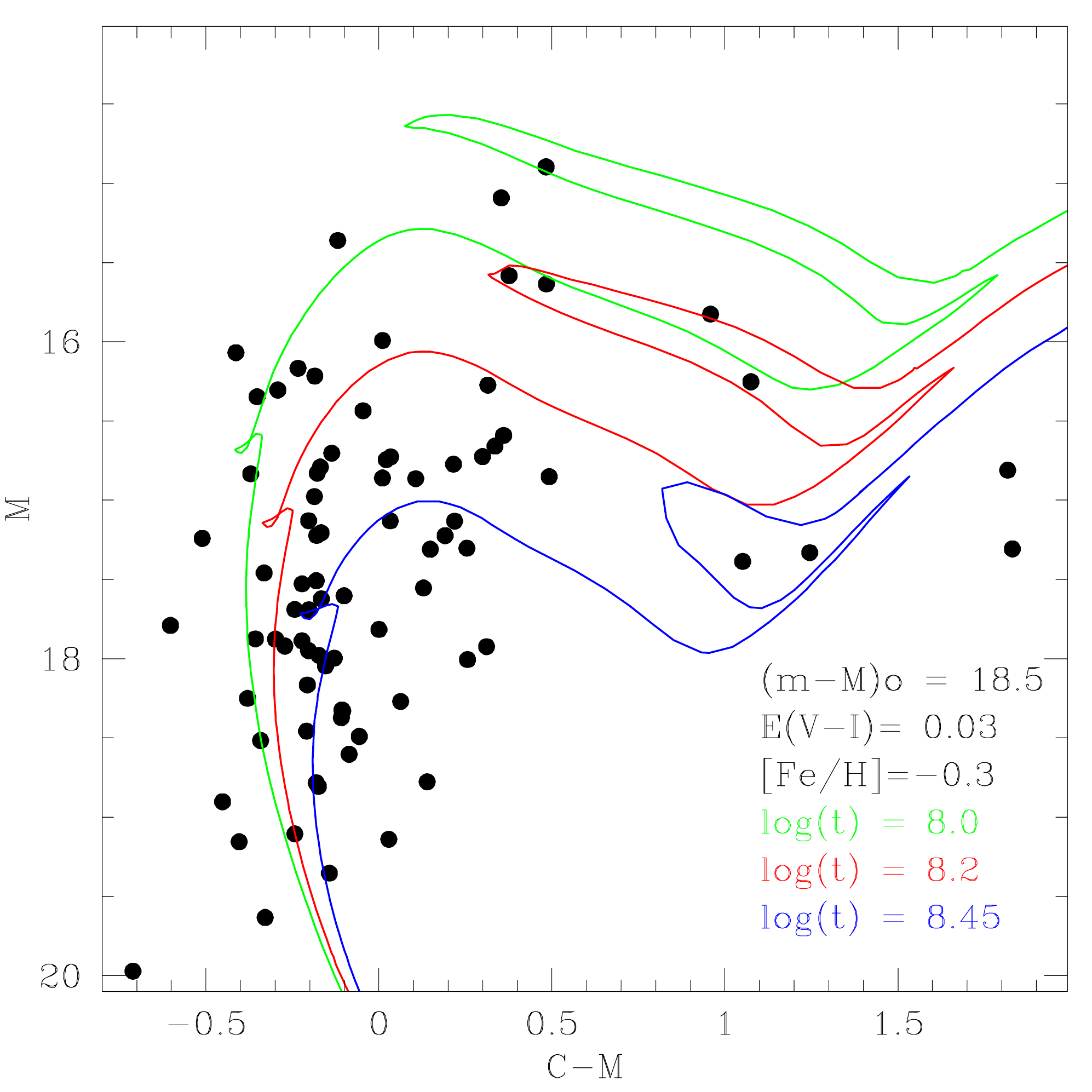}
    \caption{$M$ versus $C-M$ CMD for the observed stars in Fig.~\ref{fig:fig3}.
We also superimposed isochrones of \citet{betal12}.}
   \label{fig:fig4}
\end{figure}

\begin{figure}
	\includegraphics[width=\columnwidth]{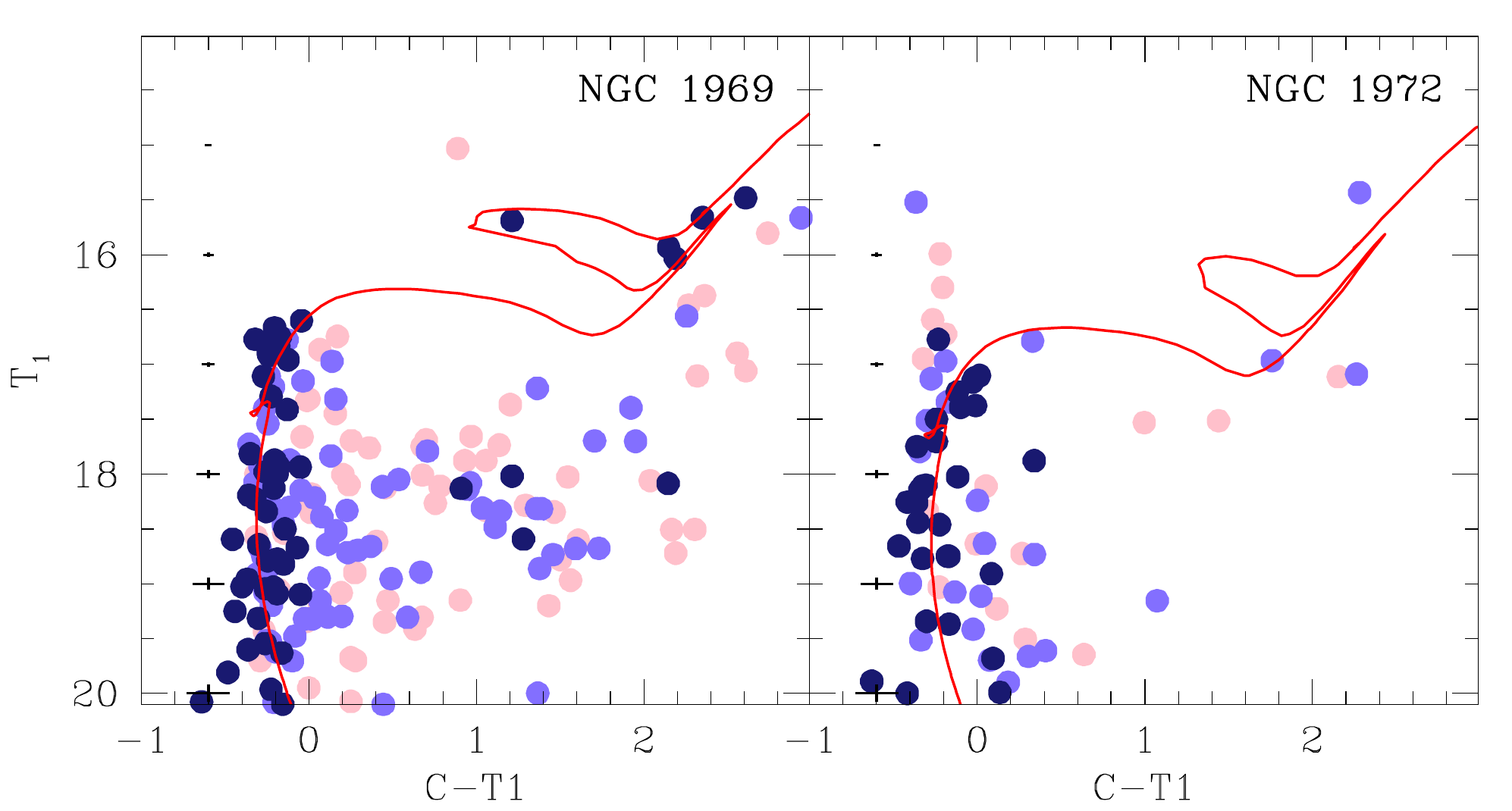}
    \caption{Cleaned CMDs for stars with $P \ge$ 75$\%$, $=$ 50$\%$ and $\le$ 25$\%$ drawn with
    dark blue, light blue and pick filled circles, respectively, located in the cluster region.
    \citet{betal12}'s isochrones are also superimposed with red lines (see text for details).}
   \label{fig:fig5}
\end{figure}

As for estimating the magnitude of stellar rotation effects, we made use of the
relationship obtained by \citet{niederhoferetal15b} between the age of a star
cluster and the expected (apparent) age spread. Thus, by interpolating
an age of 100 Myr (log($t$ yr$^{-1}$) = 8.0) in their Fig. 2, we obtained $\Delta$(age) $\sim$ 35 Myr. 
\citet{bastianetal2017} found a high fraction of Be stars 
in NGC\,1850 ($\sim$ 80 Myr) and NGC\,1856 ($\sim$ 280 Myr) that implies a high fraction of
radiply rotating stars. As H$\alpha$ contributes to the $T_1$ mag, the observed
broadness could be affected by Be stars. However, our additional Washington $M$ photometry
\citep{petal2017} confirms that the observed spread is not mainly caused by rapid
rotators (see Fig.~\ref{fig:fig4}).
Finally, we estimated the real age spread by approximating the observed and the different modelled 
(as simple stellar populations of the best fitting age) age distributions as Gaussians, and subtracted the age widths from stellar binarity and observational errors
 ($\Delta$(age) = 46 Myr), from metallicity variations ($\Delta$(age) = 20 Myr) and from 
stellar rotation from the observed width 
($\Delta$(age) = 180 Myr) in quadrature to obtain a intrinsic real age width 
of $\Delta$(age) = 170 Myr. This result reveals, for the first time, the existence
of a young LMC cluster with an eMSTO mostly originated by a real age spread. 
We also analyzed NGC\,1969 and NGC\,1972 (see Figs.~\ref{fig:fig1} and ~\ref{fig:fig5}) 
and derived ages of log($t$ yr$^{-1}$) = 8.3 and 8.4, respectively, and found no age spread
at their MSTOs.


\section*{Acknowledgements}
We thank the referee for his/her thorough reading of the manuscript and
timely suggestions to improve it. 




\bibliographystyle{mnras}

\input{paper.bbl}








\bsp	
\label{lastpage}
\end{document}